\documentclass[11pt]{article}
\usepackage{amsmath,amssymb}

\newcommand{\hE}{\hat{E}}

\newcommand{\om}{\omega}
\newcommand{\p}{\partial}

\newcommand{\hR}{\hat{R}}
\renewcommand{\a}{\alpha}
\renewcommand{\b}{\beta}

\newcommand{\reals}{\mathbb{R}}
\newcommand{\si}{\sigma}
\newcommand{\de}{\delta}
\newcommand{\tr}{\text{tr}}
\newcommand{\ap}{\alpha'}

\newcommand{\m}{{\bf m}}
\newcommand{\bn}{{\bf n}}

\newcommand{\be}{\begin{eqnarray}}
\newcommand{\ee}{\end{eqnarray}}
\newcommand{\nn}{\nonumber}

\begin{document}

\begin{center}

\begin{flushright}AEI-2007-088\\[10mm]\end{flushright}

{\Large \bf Ehlers Symmetry at the Next Derivative Order}\\[16mm]

Claudia Colonnello and Axel Kleinschmidt\\[6mm]

{\sl Max Planck Institute for Gravitational Physics, Albert Einstein
  Institute\\
Am M\"uhlenberg 1, 14476 Potsdam, Germany}\\[15mm]

\begin{abstract}

\footnotesize{\noindent
We analyse four-dimensional gravity in the presence of general
curvature squared corrections and show that Ehlers'
$SL(2,\reals)$ symmetry,
which appears in the reduction of standard gravity to three
dimensions, is preserved by the correction terms. The mechanism
allowing this is a correction of the $SL(2,\reals)$
transformation laws which resolves problems with the
different scaling behaviour of various terms occurring in the reduction.}

\end{abstract}

\end{center}

\begin{section}{Introduction}

Supergravity theories, when compactified on tori, exhibit so-called
hidden symmetries which extend the na\"ive torus symmetries
\cite{Cremmer:1978,Pope:1998,Cremmer:1999}. In  
the case of pure gravity in $(3+n)$ dimensions reduced on an
$n$-torus, there is a hidden $SL(n+1)/SO(n+1)$ symmetry which occurs
when dualising the field strengths of the Kaluza--Klein vectors
obtained in the dimensional reduction to gradients of scalar fields in
the three remaining space-time dimensions. This was
first observed in the case of $(3+1)$ dimensions a long time ago and
the enhanced symmetry $SL(2,\reals)$ is known as Ehlers symmetry
\cite{Ehlers:1957}. Many supergravity theories arise as the low
energy limit ($\ap\to 0$) of string theory. String theory predicts in
such cases also corrections to the two-derivative supergravity
actions. These corrections are a double expansion in the string
coupling $g_s$ and the squared string scale $\ap=\ell_s^2$,
potentially with non-perturbative contributions. Since
$\ap$ is dimensionful, higher powers of $\ap$ most be accompanied by
a higher number of derivatives than the supergravity
approximation, for example higher powers of the curvature
tensors. Some of these terms have been computed from string scattering
and supersymmetry, see for example
\cite{Peeters:2000,Iengo:2002,Cederwall:2004,Green:2005,Policastro:2006}.
It is interesting to study 
if the hidden symmetries of the lowest order supergravity action are
preserved by these higher order derivative corrections. 

Recently this question has been studied, with results mostly suggesting
that the hidden symmetries are broken by higher order in derivatives
corrections
\cite{Lambert:2006-1,Lambert:2006-2,Bao:2007,Michel:2007}. The purpose
of this note is to show that at least in the case of pure gravity reduced from
four to three space-time dimensions the $R^2$ extension to fourth
order in derivatives does not affect the symmetry enhancement. One
argument for the breakdown of the symmetry has been the appearance of
dilaton pre-factors in the reduced higher derivative correction terms
which seem to spoil the invariance under the enhanced symmetry
groups since dilaton scaling is one of the hidden symmetry
transformations. However, working in the context of a perturbative
$\ap$ formulation we find here that the transformation laws themselves
can be  modified at next order in  $\ap$ in 
precisely the right way to restore the invariance. The case treated 
here is very simple, and has no good string theory origin, but
can be seen as a first step towards the more complicated cases arising
in string theory and serves to illustrate a mechanism for preserving
the hidden symmetries in the presence of higher derivative terms. Our
analysis is similar in spirit to that
of~\cite{Meissner:1996sa,Kaloper:1997ux}.

This note is structured as follows. In section~\ref{trivcor} we
study pure gravity in four dimensions extended to 
fourth order in derivatives and show that in this case the most
general such correction (expressed in terms of the Riemann and Ricci 
tensors) is actually trivial in the sense that it can
be made to vanish by means of local field redefinitions. This implies
in particular that it cannot break Ehlers symmetry. We then proceed to
consider the reduction of the $R^2$ terms 
in general and exhibit explicitly a large set of field
redefinitions that effectively eliminate the higher derivative
correction and restore the explicit Ehlers symmetry of the
action. In section~\ref{scaling} we discuss the modified
$SL(2,\reals)$ action on the fields, in particular the subset
of global scaling transformations, and explain how the problem
of different scaling is resolved in this case such that the symmetry
is restored. 
Finally, in section~\ref{concl} we offer some
comments on the general case which is currently under investigation. In the
appendix we list the curvature tensors that arise in the reduction. 

\end{section}

\begin{section}{$R^2$ Corrected Action in $(3+1)$ Dimensions}
\label{trivcor}

Pure gravity in a four-dimensional space-time has a hidden
$SL(2,\reals)$ symmetry \cite{Ehlers:1957}. It is exhibited by
performing a dimensional reduction of the theory to three dimensions
and dualising the degree of freedom of the obtained Kaluza--Klein
vector to a scalar. This scalar, called the axion, together with the
dilaton obtained also from the dimensional reduction, parametrize the
coset space $SL(2,\reals)/SO(2)$. We are interested in investigating
if this hidden symmetry is affected by higher derivative
corrections.

Our starting point is then the Einstein--Hilbert action in four
dimensions and its correction by fourth order in derivative terms. The 
most general such action that one can write is\footnote{We do not
  consider the self-dual and anti-self-dual parts of the Weyl tensor
  separately but only the combination in which they appear in the
  Riemann tensor since this is the expression which generalizes to
  higher dimensions.}
\be\label{action}
\hat{S}=\int{}d^4x \hE\Big[\hR + \ap\left(\m_1\hR^{MNPQ}\hR_{MNPQ}
  +\m_2 \hR^{MN}\hR_{MN} + \m_3 \hR^2\right) \Big],
\ee
with $\{\m_i\}$ three arbitrary real constants and $\ap$ a
dimensionful parameter, with dimensions $\mbox{(Length)}^2$. Hatted
objects and capital letters denote four-dimensional quantities while
small characters are generally reserved for three-dimensional
quantities. We work in a vielbein formalism so that
$\sqrt{-\hat{G}}=\det(\hE_M{}^A)=:\hE$. 

To see that the correction to the Einstein--Hilbert action in four
dimensions is trivial, one uses field redefinitions of the metric 
and the fact that the Gauss--Bonnet combination  
\begin{equation}
\hat{R}_{GB}\,=\,\hat{R}^{MNPQ}\hat{R}_{MNPQ}
  -4\hat{R}^{MN}\hat{R}_{MN}+\hat{R}^2,
\end{equation} 
is a topological invariant in four dimensions and does not contribute
to the dynamics. This fact can be used to eliminate the correction
term proportional to the contraction of the Riemann tensor with itself
by subtracting from the action (\ref{action}) a term proportional to
the Gauss--Bonnet combination that cancels it. The remaining fourth
order contributions can be eliminated  by performing the following
field redefinition of the four-dimensional metric  
\begin{equation}\label{4Dgenred}
\hat{G}_{MN}\longrightarrow\hat{G}_{MN}+\ap\delta\hat{G}_{MN}\,
=\,\hat{G}_{MN}+\ap\left(\bn_1\hat{R}_{MN}+\bn_2\hat{G}_{MN}\hat{R}\right), 
\end{equation}
with $\bn_1=-(4\m_1+\m_2)$ and $\bn_2=(\m_1+\frac{1}{2}\m_2+\m_3)$. The
action in the redefined metric is simply the original Einstein--Hilbert
action and therefore there is a choice of fields after dimensional reduction
such that the Ehlers symmetry of (\ref{action})  is preserved.
Instead of using this field basis, we will continue considering the
action (\ref{action}) and perform explicitly its dimensional reduction to
three dimensions, as an example of the more general computation for
higher dimensions and as a tool to describing the resolution to the
problematic different scaling properties of the various orders in
$\ap$ below. In the higher-dimensional case the
Gauss--Bonnet combination is no longer a topological 
invariant and one cannot use it to eliminate a term in the
correction and the action (\ref{action}) is no longer equivalent to
the Einstein--Hilbert action. 
There are, however, simplifications to the
problem that occur in the compactification to three dimensions in the
general case. For the reduction from four to three dimensions
further simplifications arise since the local hidden symmetry $SO(2)$
is abelian.

\end{section}

\begin{section}{The Reduced Corrected Action}
\label{redaction}

We consider now the dimensional reduction of the corrected
Einstein--Hilbert action (\ref{action}) to three dimensions along a
space-like direction. The coordinates split as $x^M=(x^m,\tilde{z})$,
$M=0,..,3,\,m=0,..,2$ with $\tilde{z}$ the compact direction. Our
ansatz for the reduction of the four-dimensional metric is  
\be\label{KKansatz}
d\hat{s}^2=e^{-\phi}ds^2+e^{\phi}(d\tilde{z}+A_m dx^m)^2.
\ee
The three-dimensional metric is $ds^2=g_{mn}dx^mdx^n$ and the dilaton
dependence in this ansatz has been chosen such that the 
reduced action is in Einstein frame with standard normalization for
the terms in the lowest order reduced action. Our results are
independent of this choice but we adopt it for convenience.
The reduction of the corrected gravity
action (\ref{action}) gives the following three-dimensional action
{\allowdisplaybreaks\begin{eqnarray}\label{l}
&&\!\!\!\!\!\!\!\!
 S=\int{}d^3x e\bigg\{R - \frac{1}{2}(\partial\phi)^2
  -\frac{1}{4}e^{2\phi}{F}^2\,\\
&&\!\!\!\!\!\!+\ap\bigg[e^{\phi}\bigg((-\m_1 + \m_3)R^2
 + (4\m_1+\m_2)R^{mn}R_{mn} \nn\\
&&\!\!  - (4\m_1 + \m_2)R^{mn}\partial_m\phi\partial_n\phi
  -(2\m_1 + \m_2 + 2\m_3)D^mR\partial_m\phi\nn\\ 
&&\!\!- (\m_1 + \m_2 +3\m_3)R(\partial\phi)^2 
  - (3\m_1 + \m_2 + \m_3)\partial^m\phi\nabla_m(\square\phi)\nonumber\\  
&&\!\! -(4\m_1 + \frac{3}{2}\m_2 + 2\m_3)
          \square\phi(\partial\phi)^2 
  + \frac{1}{4}(3\m_1 + \m_2 + \m_3)(\partial\phi)^4\bigg)\nonumber\\  
&&\!\!\!\!\!+ e^{3\phi}\left(-(6\m_1 
  + \frac{3}{2}\m_2)R^{mn}F_{mp}F_n\,^p\right.
  + \frac{1}{2}(3\m_1 + \frac{1}{2}\m_2 - \m_3)R{F}^2 \nn\\
&&\!\!  - (\m_1 +\frac{1}{4}\m_2)F^{mn}\square{}F_{mn}
  - \frac{1}{2}(4\m_1 +
  \m_2)F^{mp}F^n\,_p\nabla_m\partial_n\phi \nn\\
&&\!\! + (4\m_1 +  \m_2)
       F^{mp}F^n\,_p\partial_m\phi\partial_m\phi 
   - \frac{1}{8}(16\m_1 + 5\m_2 + 4\m_3)
       {F}^2\square\phi\nn\\
&&\!\!+ \frac{1}{4}(5\m_1 +\frac{3}{2}\m_2 + \m_3)
    {F}^2(\partial\phi)^2\bigg)
  + e^{5\phi}\frac{1}{16}(11\m_1 + 3\m_2 + \m_3)({F}^2)^2\bigg]\bigg\}.\nn 
\end{eqnarray}
The curvature tensors and covariant derivatives are constructed from
the three-dimensional metric $g_{mn}$. The complete decomposition of
the four-dimensional curvature tensors 
in terms of the three-dimensional fields is given for completeness in
the appendix. To obtain the expresion (\ref{l}) we have used Bianchi
identities and the special properties of the three-dimensional 
space-time and of the abelian gauge transformations of the
Kaluza--Klein vector in addition to some integrations by parts on the
action obtained directly using the dimensional reduction ansatz
(\ref{KKansatz}). These manipulations are convenient because in
(\ref{l}) all  order $\ap$ terms 
vanish for the Gauss--Bonnet values, illustrating the fact that they
constitute a total derivative.  }

As mentioned before the lowest order terms in (\ref{l}) possesses a
hidden $SL(2,\reals)$ symmetry. To observe it one has to dualise the
Kaluza--Klein vector field strength to the gradient of a scalar field
$\chi$, the axion. This can be done by the introduction of a Lagrange
multiplier term $\int{}e\frac12\epsilon^{mnp}\p_m\chi{}F_{np}$
exchanging the Bianchi identity of the antisymmetric tensor field for
the equation of motion of the axion field
$\chi$. One can then consider $F_{mn}$ rather than $A_m$ as an
independent field and eliminate it in favour of $\chi$ by solving its
equation of motion which is only algebraic. The presence of higher
order corrections will affect the duality relation but in a
perturbative analysis one can use at each order in $\ap$ the
lower order solution. Additionally, at first order in $\ap$ the
corrections to the duality relation cancel out in the action, so that
we can safely use the lowest order relation to exchange the two
fields. The lowest order action can then be written in an explicitly
symmetric way by formulating it as  a non-linear $\sigma$-model on the
coset space $SL(2,\reals)/SO(2)$ parametrized by the dilaton and
axion. In terms of the Cartan--Killing trace it is given by 
\be\label{3Ddualaction}
\tilde{S}_0=\int{}d^3x e\left(R-\tr(P^mP_m)\right)=\int{}d^3x
e\left(R-\frac12(\p\phi)^2-\frac12 e^{-2\phi} (\p\chi)^2\right). 
\ee 
Here $P_m$ is the projection along the coset directions of the Cartan
form whose parametrization in terms of the scalar fields is given by  
\be
P_m=\left(\begin{array}{cc}\frac12\p_m\phi&\frac12e^{-\phi}\p_m\chi\\
    \frac12e^{-\phi}\p_m\chi&-\frac12\p_m\phi\end{array}\right).
\ee
The $SL(2,\reals)$ transformations only act on the scalar sector
leaving the three-dimensional metric and curvature tensors invariant. 

The first order in $\ap$ terms in (\ref{l}) seem hard to reconcile
with this symmetry. However, they can be brought into a symmetric form
by performing redefinitions on the fields, as was also done in 
\cite{Meissner:1996sa}. We consider the following general fourth order
action exhibiting explicitly Ehlers symmetry,  
\be\label{saction}
\tilde{S}_s
&=&\int{}d^3x e\left[R - \mbox{tr}(P^mP_m)\right.\nn\\
&&+\left.\ap\left(a\,\mbox{tr}(P^mP_m)^2 + b\,\mbox{tr}(P^mP^nP_mP_n) +
  cR_{c}^2\right)\right]\nn\\ 
&=&\int{}d^3x e\left\{R - \frac{1}{2}(\partial\phi)^2 -
\frac{1}{2}e^{-2\phi}(\partial\chi)^2\right.\nn\\ 
&&+ \ap\left[(a+b)\left((\partial\phi)^4 +
   2e^{-2\phi}(\partial\phi)^2(\partial\chi)^2
+ \frac{1}{4}e^{-4\phi}(\partial\chi)^4\right)\right.\nn\\
&&\left.\left.\quad\quad\quad+4be^{-2\phi}\p_a\phi\p_b\phi\p^a\chi\p^b\chi +
   cR_{c}^2\right]\right\}. 
\ee 
Here $a,b$ and $c$ are arbitrary real constants and the $R_{c}^2$ term
refers to any pure curvature fourth order in derivatives terms, the
contractions are suppressed for simplicity.\footnote{For bigger
  symmetry groups, like 
  the ones arising when compactifying gravity in more than four
  dimensions down to three, there can be four independent
  contributions of quartic traces, with $\mbox{tr}(P^mP_mP^nP_n)$
  and $\mbox{tr}(P^mP^n)\mbox{tr}(P_mP_n)$ apparently missing in
  (\ref{saction}). However, in the case of $SL(2,\reals)$ only two of
  these structures are linearly independent. We also disregard terms
  of the form $\mbox{tr}(P^mP_m)R$ or $\mbox{tr}(P^mP^n)R_{mn}$ since
  they are not needed  here and they have a different scaling
  behaviour.}  
Expressed in terms of
the graviphoton field strength and up to fourth order in derivatives
the action (\ref{saction}) becomes 
\be\label{lsymF}
S_s&=&\int{}d^3x e\left\{R - \frac{1}{2}(\partial\phi)^2 -
\frac{1}{4}e^{2\phi}{F}^2 
 + \ap\left[(a+b)\left((\partial\phi)^4 -
   e^{2\phi}(\partial\phi)^2{F}^2 \right.\right.\right.\nn\\ 
&&\left.\left.\left.\quad\quad\quad\quad+ \frac{1}{4}e^{4\phi}({F}^2)^2\right) +
   4be^{2\phi}F^{mp}F^n\,_p\p_m\phi\p_n\phi +
   cR_{c}^2\right]\right\}. 
\ee

On this action we perform the following class of field redefinitions
\begin{eqnarray}\label{redefinitions}
\delta\phi\,&=&\,a_1(\partial\phi)^2 + a_2\square\phi + a_3R +
a_4{F}^2,\nonumber\\ 
\delta{}F_{mn}\,&=&\,b_1\square{}F_{mn} +
b_2F^{l}\,_{\left[n\right.}R_{\left.m\right]l} +
b_3F^{l}\,_{\left[n\right.}\partial_{\left.m\right]}\phi\partial_l\phi
+ b_4F^{l}\,_{\left[n\right.}\nabla_{\left.m\right]}\partial_l\phi
\nonumber\\ 
&& + b_5F_{mn}(\partial\phi)^2 + b_6F_{mn}\square\phi + b_7F_{mn}R +
b_8F_{mn}{F}^2,\nonumber\\ 
\delta{}g_{mn}\,&=&\,c_1R_{mn} + c_2\partial_m\phi\partial_n\phi +
c_3\nabla_m\partial_n\phi + c_4F_{mp}F_{n}\,^p + \nn\\
&&+g_{mn}\left(d_1(\partial\phi)^2 + d_2\square\phi + d_3R + d_4{F}^2\right). 
\end{eqnarray}
The coefficients $\{a_i, b_i, c_i, d_i\}$  are real parameters and may
have a dilaton dependence in the form of an exponential
prefactor. These are the most general field redefinitions that produce
terms in the action like the ones appearing in the reduced higher
order action (\ref{l}). They are such that the axion $\chi$ has a
dual field strength which is local in the original fields. 

Under these redefinitions the symmetric action (\ref{lsymF}) gets
modified at non-zero orders in $\ap$ to become, to first order 
\begin{eqnarray}\label{redl}
S_s\,&=& S_0 + \ap\int{}d^3x e\left\{(a+b)\left((\partial\phi)^4 -
e^{2\phi}(\partial\phi)^2{F}^2\right.\right.\nn\\  
&&\left.\left.+ \frac{1}{4}e^{4\phi}(F^2)^2\right) +
4be^{2\phi}F^{mp}F^n\,_p\p_m\phi\p_n\phi + cR_c^2 \right.\nn\\
&&- \frac{1}{2}e^{2\phi}F^{mn}\delta{}F_{mn}
-\left(\frac{1}{2}e^{2\phi}{F}^2 - \square\phi\right)\delta\phi\nn\\
&&-\delta{}g_{mn}\left[ R^{mn} - \frac{1}{2}g^{mn}\left(R -
  \frac{1}{2}(\partial\phi)^2 -
  \frac{1}{4}e^{2\phi}{F}^2\right)\right.\nonumber\\ 
&&-\left.\left.\frac{1}{2}\partial^m\phi\partial^n\phi -
  \frac{1}{2}e^{2\phi}F^{mp}F^{n}\,_p 
\right]\right\}.
\end{eqnarray}

Inserting the redefinitions (\ref{redefinitions}) and performing some
integrations by parts brings (\ref{lsymF}) into the form of the action
(\ref{l}) allowing one to compare the coefficients in both expressions
term by term, which results in a system of linear equations for the
arbitrary coefficients in (\ref{redefinitions}) in terms of $\{\phi,
\m_1, \m_2, \m_3 \}$. The system turns out to be solvable for
arbitrary values of $\m_1, \m_2$ and $\m_3$, as was expected  from the
general considerations of the previous section. This is sufficient to
show that Ehlers symmetry is not broken by a general order $\ap$
correction to the Einstein--Hilbert action.

The actual solution is not unique, there turns out to be some
arbitrariness in the choice of some of the coefficients. The general
result is 
{\allowdisplaybreaks\be\label{fixedredefinitions}
\delta\phi\,&=&\,\left[\frac{1}{2}c_3+\frac{1}{4}d_2
    -\frac{1}{2}e^{\phi}(2\m_1+\m_2+2\m_3)\right](\partial\phi)^2\nn\\ 
&& +e^{\phi}\left(3\m_1+\m_2+\m_3\right)\square\phi \nonumber\\  
&& + \left[-\frac{1}{2}d_2+e^{\phi}(2\m_1+\m_2+2\m_3)\right]R 
   +a_4{F}^2,\nonumber\\ 
\delta{}F_{mn}\,&=&\,\frac{1}{2}e^{\phi}(4\m_1+\m_2)\square{}F_{mn}\nn\\ 
&&+ e^{-2\phi}\left[-2c_4+2e^{3\phi}(4\m_1+\m_2)\right]
    F^{l}\,_{\left[n\right.}R_{\left.m\right]l}\nonumber\\   
&&+  e^{-2\phi}\left[e^{2\phi}c_3+c_4+8e^{2\phi}b
   -\frac{3}{2}e^{3\phi}(4\m_1+\m_2)\right]  F^{l}\,_{\left[n\right.}
           \partial_{\left.m\right]}\phi\partial_l\phi\nonumber\\  
&&+ \left[c_3+e^{\phi}(4\m_1+\m_2)\right]F^{l}\,_{\left[n\right.}
   \nabla_{\left.m\right]}\partial_l\phi \nonumber\\ 
&&+ e^{-2\phi}\left[-\frac{3}{4}e^{2\phi}c_3-\frac{1}{4}e^{2\phi}d_2
      -\frac{1}{2}c_4-\frac{1}{2}d_4\right.\nn\\ 
&&\left.\quad-\frac{1}{4}e^{3\phi}(9\m_1+2\m_2-\m_3)\right]
   F_{mn}(\partial\phi)^2 \nonumber\\   
&&+ \frac14\left[-c_3+d_2+8e^{-2\phi}a_4
    +e^{\phi}(4\m_1+\m_2)\right]F_{mn}\square\phi \nonumber\\   
&&+ \frac12e^{-2\phi}\left[e^{2\phi}d_2+2c_4+2d_4
    -e^{3\phi}(7\m_1+2\m_2+\m_3)\right]F_{mn}R \nonumber\\
&&+ \frac14\left[-4a_4+c_4+d_4
    -\frac{1}{2}e^{3\phi}(11\m_1+3\m_2+\m_3)\right]F_{mn}{F}^2,\nonumber\\
\delta{}g_{mn}\,&=&\,-e^{\phi}(4\m_1+\m_2)R_{mn} 
+ \left[c_3+\frac{1}{2}e^{\phi}(4\m_1+\m_2)\right]\partial_m\phi\partial_n\phi\nn\\
&&+ c_3D_m\partial_n\phi+ c_4F_{mp}F_n\,^p\nonumber\\
&&+ g_{mn}\,\left[-\frac{1}{2}e^{\phi}(2\m_1+\m_2+2\m_3)(\partial\phi)^2 
+ d_2\square\phi\right.\nn\\ 
&&\left.\quad+ e^{\phi}(2\m_1+\m_2+2\m_3)R 
+ d_4{F}^2\right],\nn\\
&&\mbox{}\nn\\
a&=&-b,\hspace{1.0cm}c=0. 
\ee
The  six parameters ${(b,a_4,d_2,d_4,c_3,c_4)}$ are free, the solution
is highly degenerate. For consistency one expects that the particular
field redefinition (\ref{4Dgenred}) which eliminated
the whole correction term at the four-dimensional level should be an
allowed solution. In fact, one
can compute the redefinitions on the three-dimensional fields induced
by (\ref{4Dgenred}) after performing the dimensional reduction and
show that these correspond to (\ref{fixedredefinitions}) for a
particular choice of the six free parameters. For this choice the
correction $\de F_{mn}$ can be traced back to a correction to the
Kaluza--Klein vector potential $A_m$. }

Now that we have proved that the fourth order correction does not
affect the hidden symmetry of the action, we will describe in some
more detail the action of a subset of the $SL(2,\reals)$
transformations and show how they are modified by fourth order terms. 

\end{section} 

\begin{section}{Corrected  $SL(2,\reals)$ Transformations}  
\label{scaling}

At lowest order the Ehlers $SL(2,\reals)$ transformations include
scaling transformations of the fields, $\phi \longrightarrow\phi +
\si$, $\chi \longrightarrow e^\si\chi $, associated with the Cartan
generator of $SL(2,\reals)$. One can then read off the transformation
of the four-dimensional curvature objects from their expresions in terms of the
lower-dimensional fields obtained when performing the dimensional
reduction. One finds that the scalar curvature and four-dimensional
measure transforms under this lowest order global rescaling as
(cf. formulas in the appendix)
\be\label{zeroscalings} 
\hR \longrightarrow e^{\si} \hR,\quad\hE = e^{-\phi} e \longrightarrow
e^{-\si} \hE. 
\ee
Clearly the combination $\hE \hR$ is then invariant, however, the
higher order contribution $\hE \hR^2$ is not 
invariant and therefore seems to break the $SL(2,\reals)$
symmetry. This presents an apparent contradiction with our previous
results where we showed that the symmetry is preserved by general
$R^2$ order curvature corrections and, moreover, that these are
trivial in this case.  

The problem arises from the appearance of overall exponential
prefactors of the dilaton carried by the expansion of the higher
dimensional curvature objects after dimensional reduction, as is
evident from the explicit results in the appendix. At lowest
order these prefactors can be made to cancel out with the ones coming
from the reduction of the measure by choosing an appropriate frame,
most conveniently the Einstein frame as in
(\ref{zeroscalings}). However, then they will not cancel 
for the higher order curvature terms. This is a 
generic feature of the dimensional reduction of supergravity actions
and is related to the observation in
\cite{Damour:2005,Lambert:2006-1,Damour:2006ez} according 
to which higher curvature terms are associated with the weights of the
hidden symmetry algebra rather than its roots which occur in the usual
parametrization of the coset space. The dilaton prefactors are at the
same time associated to the volume of the compactified space and are
central when trying to complete the weight structure by
an automorphic form of the discrete version of the hidden symmetry
algebra \cite{Lambert:2006-2,Bao:2007,Michel:2007}.
 
However, in this simple case the symmetry as we have seen is not
broken. The solution to the puzzle is that the transformation
rules for the original $\phi$, $\chi$ and the metric $g_{mn}$ are only
correct up to lowest order and change once $\ap$ terms are considered,
giving new 
contributions to the transformation of the action that restore the
symmetry. To derive the modification of the transformation rules of
the three-dimensional fields one needs to notice that the
redefined fields, in terms of which the symmetry of the action is
manifest, transform by construction just under the usual (lowest
order) Ehlers transformations. The original fields then have an
expansion in terms of the redefined ones. From this one can derive the
$\ap$ corrections to the transformation laws 
of the original fields. The result in the case of the scaling
transformation is the following,   
\be\label{finiteredscaling}
\phi &\longrightarrow& \phi + \si + \ap\delta\phi(1-e^{\si}),\nn\\
F_{mn} &\longrightarrow& e^{-\si}F_{mn} + \ap\delta{}F_{mn}(e^{-\si}-1),\nn\\
g_{mn} &\longrightarrow& g_{mn} + \ap\delta{}g_{mn}(1-e^{\si}).
\ee{}
It is easy to see that, when acting on the lowest order
Einstein-Hilbert action, the $\ap$ terms in the transformations
(\ref{finiteredscaling}) generate a term that reconstructs the
original correction to the action globally rescaled by a factor of
$e^{\sigma\phi}$, which precisely cancels the actual correction terms
transformed at lowest order as follows from (\ref{zeroscalings}). The
symmetry is then restored up to fourth order in derivatives. The other
transformations that make up the $SL(2,\reals)$ group can be analysed
in a similar way and order $\ap$ modifications to them also appear. 

\end{section}

\begin{section}{Outlook}
\label{concl}

The case of four-dimensional gravity reduced to three dimensions
studied in this paper contains a number of non-generic features.
It is not clear to what extent the result that the hidden symmetry
is unbroken by general $R^2$ corrections carries over to the
reduction of $R^2$ extended $(3+n)$-dimensional gravity to three
dimensions. Nevertheless we believe that the techniques used in
this paper can be employed to address this question and our
results emphasise that $\ap$-corrections to the symmetry
transformations are an important feature to be considered.

One of the special properties of $R^2$ corrections in $(3+1)$
dimensions is the topological nature of the Gauss--Bonnet
combination which can be used to remove the square of the full
Riemann tensor. The remaining terms are then proportional to the lowest
order equations of motion and can be removed by field
redefinitions in $(3+1)$ dimensions.
This clearly is no longer possible in the general
case since $\hR_{ABCD}\hR^{ABCD}$ is not part of a total derivative.
Furthermore, one cannot generate such a term from a field
redefinition in $(3+n)$ dimensions before reduction for $n>1$.
We still deem it likely that this term can be accounted for in a
symmetric action by
a field redefinition {\em after} reduction to three
dimensions since the Weyl tensor vanishes identically and the
three-dimensional Riemann tensor therefore can be expressed in terms
of the Ricci tensor and scalar, both of which do arise from field
redefinitions. First computations indicate that this can be
implemented and that also the other terms arising in the
reduction can be treated in a similar fashion. We plan to report
on this in more detail in the near future.

String effects and in particular string loops are expected to break
the hidden symmetry arising in dimensional reduction to its arithmetic
version \cite{HuTo,ObPi}. This is consistent with the present 
analysis where we only considered perturbative $\ap$ corrections to
gravity. In the full theory it seems likely that automorphic forms are
needed in order to maintain the discrete symmetry.

\end{section}

\vspace{5mm}
{\bf Acknowledgements}\\
We are grateful to L. Bao, M. Cederwall, K. Meissner, H. Nicolai, B.E.W.
Nilsson, K. Peeters and J. Plefka for discussions. AK would like
to thank Chalmers University, Gothenburg,
 for its warm hospitality and support. We thank Y.Michel and B.Pioline for correspondence.

\appendix

\begin{section}{Reduction of Curvature Tensors}

Here, for completeness, we list the results of the dimensional
reduction of the curvature tensors from four to three dimensions
derived from the general ansatz 
\be\label{genansatz}
d\hat{s}^2=e^{2\a\phi}ds^2+e^{2\b\phi}(d\tilde{z}+A_mdx^m)^2.
\ee  
Again, hatted objects and capital letters denote four-dimensional
quantities while small characters are reserved for three-dimensional
ones. The coordinates split as $x^M=(x^m,\tilde{z}),\;
M=0,..,3,\,m=0,..,2$ with $\tilde{z}$ the space-like compact
direction. Flat indices are denoted by letters at the beginning of the
alphabet and split as $A=(a,z),\; A=0,..,3,\,a=0,..,2$. They are
contracted using the flat Minkowski metric
$\hat{\eta}_{AB}=\mbox{diag}(-,+,+,...)$. In (\ref{genansatz}) $\a,\b$
are arbitrary real constants, in general we choose them to have values
$\a=-\b=-\frac12$ in order for the the lowest order reduced gravity
action to be in the Einstein frame. The tangent space components of
the curvature tensors then are
{\allowdisplaybreaks\be\label{riemannred}
\hR_{azbz} &=& e^{-2\a\phi}\left[-\b D_a\p_b\phi
 + \b(2\a-\b) \p_a\phi \p_b\phi \right.\nn\\
&&\quad\quad\quad\left.-\a\b \eta_{ab} (\p\phi)^2  +\frac14
 e^{(-2\a+2\b)\phi} F_{ac}F_{bc}\right],\nn\\ 
\hR_{azbc} &=& e^{(-3\a+\b)\phi}\left[-\frac12D_a F_{bc}
 +\a\p_e\phi F_{e[b}\eta_{c]a} \right.\nn\\
&&\left.\quad\quad\quad\quad\quad +(\a-\b)\p_a\phi
 F_{bc}-(\a-\b)\p_{[b}\phi F_{c]a}\right],\nn\\ 
\hR_{abcd} &=& e^{-2\a\phi}\bigg[ R_{abcd}
  -2\a D_a \p_{[c}\phi \eta_{d]b}
  +2\a D_b \p_{[c}\phi \eta_{d]a}\nn\\
&&\quad\quad\quad  +2\a^2 \p_a\phi \p_{[c}\phi \eta_{d]b} -2\a^2
  \p_b\phi \p_{[c}\phi \eta_{d]a} -2\a^2 (\p\phi)^2
  \eta_{a[c}\eta_{d]b}\nn\\ 
&&\quad\quad\quad   -\frac12e^{(-2\a+2\b)\phi}F_{ab}F_{cd}
  +\frac12e^{(-2\a+2\b)\phi}F_{a[c}F_{d]b}\bigg],\nn\\ 
\hR_{zz} &=& e^{-2\a\phi}\bigg[-\b\square\phi
  -\left(\b^2+\a\b\right)(\p\phi)^2
  +\frac14 e^{(-2\a-2\b)\phi} F^2\bigg],\nn\\
\hR_{za} &=& -\frac12 e^{(-3\a+\b)\phi}\bigg[
  D^c F_{ca} + \left(-\a+3\b\right)\p_c\phi F_{ca}\bigg],\nn\\
\hR_{ab} &=& e^{-2\a\phi}\bigg[R_{ab} - \left(\a+\b\right)D_a\p_b\phi\nn\\
&&\quad\quad\quad+\left(2\a\b-\b^2+\a^2\right)\p_a\phi\p_b\phi
  -\a\eta_{ab}\square\phi\nn\\ 
&&\quad\quad\quad-\left(\a^2+\a\b\right)\eta_{ab} (\p\phi)^2 -\frac12
  e^{(-2\a+2\b)\phi} F_{ac}F_{bc}\bigg],\nn\\ 
\hR &=&e^{-2\a\phi}\bigg[R- 2\left(2\a+\b\right)\square\phi
  -\frac14e^{(-2\a+2\b)\phi}F^2\nn\\
&&\quad\quad\quad  -2\left(\a\b+\a^2+\b^2\right)(\p\phi)^2\bigg].
\ee
Here, $D_a$ is the $SO(1,2)$ Lorentz covariant derivative in three
dimensions, e.g. $D_a\p_b\phi = \p_a\p_b\phi +
\om_{ab}{}^c\p_c\phi$. A good review on the subject
of dimensional reduction can be found in \cite{Pope:KK}.} 

\end{section}

\setlength{\baselineskip}{11pt}

\end{document}